\documentclass[]{jpconf} 
\usepackage{amssymb}
\usepackage{amsmath}
\newcommand{\be}{\begin{equation}}
\newcommand{\ee}{\end{equation}}
\newcommand{\bea}{\begin{eqnarray}}
\newcommand{\eea}{\end{eqnarray}}
\newcommand{\bitem}{\begin{itemize}}
\newcommand{\eitem}{\end{itemize}}

\newcommand{\benum}{\begin{enumerate}}
\newcommand{\eenum}{\end{enumerate}}
\newcommand{\bc}{\begin{center}}
\newcommand{\ec}{\end{center}}
\begin{document}
\title{Gravitational vacuum energy in our recently accelerating universe}
\address{Departamento de Astronomia, Universidad de Chile, Santiago, Chile}
\author{Sidney Bludman}
\ead{sbludman@das.uchile.cl}
\begin{abstract}
We review current observations of the homogeneous cosmological expansion which, because they measure only kinematic variables, cannot determine the dynamics driving the recent accelerated expansion. The minimal fit to the data, the flat $\Lambda CDM$ model, consisting of cold dark matter and a cosmological constant, interprets $4\Lambda$ geometrically as a classical spacetime curvature constant of nature, avoiding any reference to quantum vacuum energy. (The observed Uehling and Casimir effects measure forces due to QED vacuum polarization, but not any  quantum material vacuum energies.)  An Extended Anthropic Principle, that Dark Energy and Dark Gravity be indistinguishable, selects out flat $\Lambda CDM$.  Prospective cosmic shear and galaxy clustering observations of the growth of fluctuations are intended to test whether the 'dark energy' driving the recent cosmological acceleration is static or moderately dynamic.  Even if dynamic, observational differences  between an additional negative-pressure material component within general relativity (Dark Energy) and low-curvature modifications of general relativity (Dark Gravity) will be extremely small.
\end{abstract}
\section{Homogeneous expansion of our universe has recently accelerated}
CMB, baryon oscillation, supernova and weak lensing
cosmographic measurements of the homogeneous expansion history of our
universe are summarized in consistent with either a static
finely-tuned cosmological constant or a dynamic mechanism,
which may be material Dark Energy or low-curvature modifications of Einstein gravity (Dark
Gravity). The homogeneous expansion measures only kinematic variables, in particular the evolution of the cosmological deceleration $q:=-a \ddot{a}/\dot{H}^2=d H^{-1}/dt-1$, which can be given a cosmological fluid description:  If the expansion rate $H^2:=(8 \pi G_N/3)\rho:=\rho/3 M_P^2$, is used to define a cosmological 'mass density', 'enthalpy density', 'pressure' and 'equation of state' are also defined:
\be \rho+P:=-2M_P^2 \dot{H},\qquad P:=-M_P^2 (3H^2+2\dot{H}),\qquad w:=P/\rho=-(3 H^2+2\dot{H})/3 H^2. \ee
In these definitions, $\rho:=3 M_P^2 H^2$ can be
separated into the ordinary cold dark matter component $\rho_m$ and a 'dark energy' $\rho_{DE}:=\rho-\rho_m$, which may be a new negative-pressure material Dark Energy or a Dark Gravity modification to Einstein general relativity.

The observed expansion history then fixes either a fine-tuned scalar field Dark Energy potential or a fine-tuned Dark Gravity modification to the GR Friedmann expansion rate \cite{Blud}.  The
observations of the homogeneous expansion history $H^2 (t)$ do not distinguish between Dark Energy and Dark Gravity and cannot fix the underlying dynamics driving the recent acceleration since redshift $z \lesssim 0.46$.
\section{Concordance model: cosmological constant as a classical constant of nature}
The minimal model fit to the observed 5-year WMAP CBR, Type Ia supernova, and Baryon Acoustic Oscillation galaxy distribution data is the flat $\Lambda CDM$ model consisting of cold dark matter and a cosmological constant
\cite{Spergel}:
\be \Omega_b,\quad \Omega_c, \quad \Omega_{\Lambda}=\qquad 0.0462\pm 0.00115, \quad 0.233\pm 0.013, \qquad 0.721\pm 0.015 , \ee
in which the $\Omega_i$ are the present baryon, cold dark matter, and cosmological constant or 'Dark Energy' densities, each measured in units $\rho_{cr}:=3 H_0^2 M_P^2$.
This one-parameter fit is fully consistent with constant $\Lambda$ or Dark Energy enthalpy $(\rho+P)/\rho=1+w=0\pm 10\%$, or 'equation of state' $w=-1$.

The simplest interpretation of constant $\Lambda$, due to Einstein, is as a classical spacetime curvature $4\Lambda$, giving {\em geometric structure} to empty spacetime. The alternative interpretation, due to Zeldovich, interprets constant $\Lambda$ as a uniform gravitational vacuum energy density, and brings forth the Cosmological Constant Problem: why quantum matter vacuum energy fluctuations apparently do not gravitate. (The Uehling and Casimir
{\em forces} on electric charges or conducting, semi-conducting or dielectric materials show only the polarization of the QED vacuum, but not the QED vacuum {\emph energy} itself \cite{Jaffe}). The Cosmological Constant Problem is a
{\em geometric} problem, only for ambitious theorists who want to calculate $\Lambda$ \cite{BludRud}.  So far, this attempt to calculate $\Lambda$ along with other constants of nature has been fruitless. Pragmatically, there is no problem if $\Lambda$ is accepted as a constant of nature like the observed electron mass or fine-structure constant.
\section{Observers present existence selects among values of $\Lambda$}
There still remains the  Cosmological Coincidence Problem: Why do we live at so late a time that the material energy density has diluted down to a comparable value $\sim \Lambda/2$?  This is a {\em material} problem, calling for an understanding of the observers' role in cosmology.  An Enhanced Weak Anthropic Principle, that observers exist now but cannot distinguish between Dark Energy and Dark Gravity, would select out the Concordance Model $\Lambda CDM$, consistent with current observations and predicting that there will be no dynamical effects on the growth of fluctuations.

\section{Only growth of fluctuations can resolve cosmodynamic degeneracy}
The challenge therefore is to show whether the Cosmological 'Constant' is truly static or dynamic, either by showing $w\neq -1$ or
by observing the growth of cosmological fluctuations.  We already know \cite{Spergel} that,
for constant $w(z)$
\be -0.11 \leq w \leq 0.14 (95\% CL),\qquad \mbox{present spatial curvature} -0.0175 \leq                           \Omega_k \leq 0.0085 (95\% CL),\ee
with weaker constraints on time-varying Dark Energy.
Improved measurements of the homogeneous expansion can never prove $w=-1$ exactly.
Because the 'dark energy' evolution at low red-shifts is at most quasi-static, any dynamical effects on the growth of fluctuations will be minimal, unless $w(z)$ evolves strongly at high red-shifts $z>5$. Difficult as they will be, projected
observations of galaxy clustering and of gravitational weak gravitational lensing cosmic shear
are needed to test $\Lambda CDM$ by potentially distinguishing static from dynamic `dark
energy', and dynamic Dark Energy from Dark Gravity.
\section*{Acknowledgement}
SAB acknowledges support from the Millennium Center for
Supernova Science through grant P06-045-F funded by Programa
Bicentenario de Ciencia y Tecnolog\'ia de CONICYT and Programa
Iniciativa Cient\'ifica Milenio de MIDEPLAN.
\section*{References}
\bibliographystyle{iopart-num}
\bibliography{bibliographyPhysrev}
\end{document}